# Fast prototyping of an SDR WLAN 802.11b receiver for an indoor positioning system

Erick Schmidt, *The University of Texas at San Antonio*
David Akopian, *The University of Texas at San Antonio*

**BIOGRAPHIES**

**Erick Schmidt** received his B.S. degree from Monterrey Institute of Technology and Higher Education, Mexico, in 2011 and his M.S. degree from The University of Texas at San Antonio (UTSA), USA, in 2015. He is currently pursuing his Ph.D. degree at UTSA. His research interests include software-defined radio for fast prototyping, WLAN indoor localization, and interference mitigation techniques for Global Navigation Satellite System (GNSS).

**David Akopian** is a Professor at the University of Texas at San Antonio (UTSA). Dr. Akopian received his Ph.D. degree from Tampere University of Technology, Finland, in 1997. Dr. Akopian's current research interests include signal processing algorithms for communication and navigation receivers, and implementation platforms for software-defined radio, and mHealth. Dr. Akopian is a Senior Member of IEEE, member of the Institute of Navigation (ION), and Fellow of National Academy of Inventors.

**ABSTRACT**

Indoor positioning systems (IPS) are emerging technologies due to an increasing popularity and demand in location based service (LBS). Because traditional positioning systems such as GPS are limited to outdoor applications, many IPS have been proposed in literature. WLAN-based IPS are the most promising due to its proven accuracy and infrastructure deployment. Several WLAN-based IPS have been proposed in the past, from which the best results have been shown by so-called fingerprint-based systems. This paper proposes an indoor positioning system which extends traditional WLAN fingerprinting by using received signal strength (RSS) measurements along with channel estimates as an effort to improve classification accuracy for scenarios with a low number of Access Points (APs). The channel estimates aim to characterize complex indoor environments making it a unique "signature" for fingerprinting-based IPS and therefore improving pattern recognition in radio-maps. Since commercial WLAN cards offer limited measurement information, software-defined radio (SDR) as an emerging trend for fast prototyping and research integration is chosen as the best cost-effective option to extract channel estimates. Therefore, this paper first proposes an 802.11b WLAN SDR beacon receiver capable of measuring RSS and channel estimates. The SDR is designed using LabVIEW (LV) environment and leverages several inherent platform acceleration features that achieve real-time capturing. The receiver achieves a fast-rate measurement capture of 9 packets per second per AP. The classification of the propose IPS uses a support vector machine (SVM) for offline training and online navigation. Several tests are conducted in a cluttered indoor environment with a single AP in 802.11b legacy mode. Finally, navigation accuracy results are discussed.

**1 INTRODUCTION**

Pedestrian navigation has become ubiquitous and evidently important in today's pervasive mobile computing world. An example of a popular navigation system is GPS [1], which uses a trilateration principle but has the limitation of working only in outdoor environments and typically requires line of sight (LOS), thus having a low performance in urban canyons, tunnels and other GPS-denied areas. Because of this, the need for an indoor positioning system (IPS) is desired. There are many WLAN-based IPS methods proposed recently in literature that use packet information such as using time-of-arrival (TOA), time-difference-of-arrival (TDOA), angle-of-arrival (AOA), and other hybrid methods which use a form of triangulation to estimate an indoor location [2]-[5]. A drawback from said techniques is a requirement for LOS and/or extra sensors and antennas. There are also other IPS solutions which rely on techniques that are not based in LOS but require additional hardware for deployment [6]. Popular IPS approaches based on WLAN signals have been used recently for their readily deployed and available infrastructure. Said WLAN-based IPS use the received signal strength (RSS) from several WLAN access points (APs) to determine or classify a location without LOS or extra sensors, i.e., by simply collecting so-called WLAN beacon frame packets associated to a RSS and an AP. Such solutions are namely fingerprinting-based IPS [7], [8].

WLAN fingerprinting IPS deployment and operation is divided into two phases: offline and online phase. The offline phase consists of collecting RSS measurements from WLAN access points (APs) to associate to known indoor locations in a grid. A database is then generated with said RSS measurements and grid locations to form unique signatures for each location in a so-called "radio-map". Afterwards, the online phase consists of a mobile device collecting RSS measurements to predict user location by associating said measurements with the previously generated radio-map via classification, for actual user navigation.

RSS obtained from WLAN infrastructure is the most common variable used to assign unique fingerprints on each location because of the availability of said measurements from commercial WLAN cards. In a way, the RSS measurements that form the radio-map describe the indoor environment for user navigation. Similarly, in an attempt to better describe the radio-map by improving the uniqueness of the fingerprints, another potential variable is proposed: the channel estimate. Typically, indoor areas are complex to describe, thus introducing multipath and other fading phenomena. By estimating the channel in an indoor environment, and therefore the fading phenomena information of each indoor location, the channel estimation variable could prove to be a unique signature for improved classification in online navigation phase. Therefore, this paper examines an IPS by obtaining RSS and channel estimate measurements from WLAN-based systems to attain a better description of the radio-map, and therefore, improved user position accuracy.

Although channel estimation coefficients are usually not obtained directly from commercial WLAN cards, there are other ways to extract these measurements. By utilizing technologies of radio receiver implementation by means of software, software-defined radio (SDR) becomes an approach. Recently, research has increases the performance of SDR receivers, as new proper instrumentation and software is released to support these efforts. Therefore, SDR solutions become popular because of providing full control of receiver baseband modules, so the researchers can integrate and test their methods without redesigning all receiver chains. This becomes an advantage for SDR based research for fast prototyping [9], [10]. By combining novel software-defined radio (SDR) solutions and fast prototyping platforms such as LV, a WLAN SDR receiver becomes a feasible choice for the proposed research. With proper SDR, the channel estimate can be extracted from baseband modules of the receiver via realistic and less time-expensive implementations methods. If proven useful, the location accuracy should improve even in scenarios with low number of APs. This paper leverages the SDR prototyping platform LV for a WLAN receiver implementation by exploiting acceleration factors [9]-[11].

As for classification, research has proposed several IPS based on methods such as k-nearest neighbor (KNN) [12], Bayesian approaches [12], [13], and support vector machines (SVM) [13]. Previous suggestions from [13] have proven SVM to have a good accuracy when compared to KNN and Bayesian approaches. Additionally, this paper provides a further study based on previous work in [14] which have proven SVM to be a good approach for accuracy. Eventually, the usefulness of indoor location based on the presence of low APs and considering channel estimation apart from the conventional RSS measurements is feasible with a proper instrumentation SDR receiver. This paper discusses an SDR implementation of an 802.11b receiver that collects WLAN beacon frame packets to obtain RSS and channel estimate measurements that are otherwise not available from commercial WLAN cards. The described SDR receiver achieves real-time packet collection by use of acceleration features from LV platform, and finally it tests an IPS based on fingerprinting techniques with an SVM classification solution to study indoor location with low APs relying on RSS and channel estimates. The next section describes overall beacon frame packets from the WLAN 802.11b standard and how these packets are extracted for the SDR receiver.

**2 WLAN 802.11B BEACON FRAMES**

The IEEE 802.11 standard has several releases which rely on slightly different modulation formats and architectures, e.g., a/b/g/n/ac [15]. Based on the standard, a mandatory requirement of said releases is to always be backwards compatible with its previous releases, and eventually to the 802.11b standard. The 802.11b release uses channels that are 22 MHz wide and also uses direct sequence spread-spectrum (DSSS) modulation, as opposed to 802.11a/g/n which relies on orthogonal frequency-division multiplexing (OFDM) modulation and has 20 MHz wide channels. Because of said compatibility, this paper addresses the 802.11b standard with DSSS modulation. Figure 1 shows DSSS channels for the 2.4 GHz band.

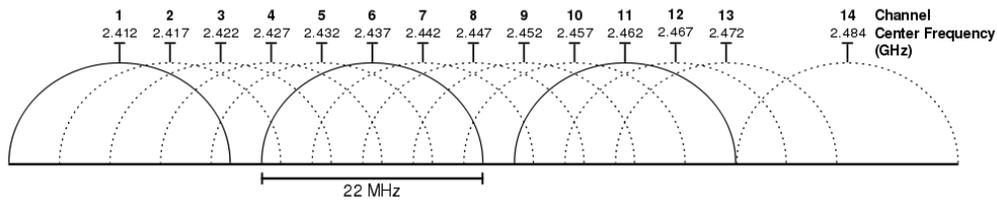

**Figure 1. 802.11b DSSS channels in the 2.4 GHz band.**

There are several types of packets being transmitted by APs and mobile devices, namely: data, management, and control packets [16]. Specifically, APs are constantly broadcasting certain management packets so that their network can be found, and for other purposes. In simple terms, a specific management packet called a beacon frame is the one which contains the AP broadcast information and which is used by mobile devices to read RSS measurements to each AP. This packet is of our interest for our proposed IPS and at the same time simplifies our receiver implementation by exclusively collecting said frames and avoid other packet types. Additionally, there is no requirement to be "connected" to a specific AP to be able to read their beacon frame since these packets are being broadcast and any mobile device can "listen" to them [16]. Therefore, we explain the beacon frame structure for 802.11b DSSS signals in the following and as seen in Figure 2.

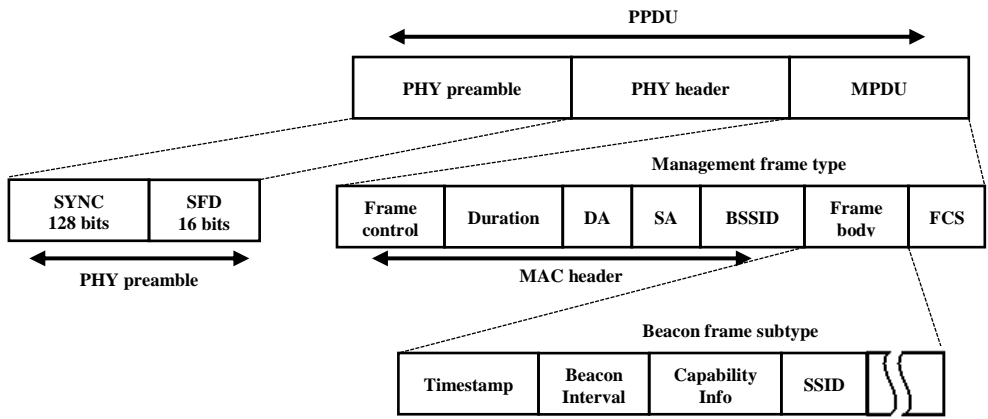

**Figure 2. Physical layer packet data unit (PPDU) structure for 802.11b beacon frames [16].**

The content of captured beacon frame packets is called the physical (PHY) layer packet data units (PPDU). The PPDU is typically divided into PHY portion which corresponds to physical layer and contains such lower layer information of the packet, and a medium access control (MAC) portion, which contains higher layer level information such as the SSID. The PHY layer is further divided into a preamble and a header as seen in Figure 2. The PHY preamble contains a training sequence called the SYNC, which consists of a known sequence of bits and is used for packet detection and synchronization. Additionally, this PHY preamble is also used for channel equalization and therefore channel estimation, as will be described later in the SDR implementation section. Once the PHY preamble is detected, a potential WLAN packet is found. The PHY header contains relevant information such as modulation type, and total packet length that is used to demodulate the rest of the PPDU, namely the MAC packet data unit (MPDU) where the rest of the relevant beacon frame information resides. A WLAN receiver will report an RSS measurement every time it successfully decodes the full beacon frame. This is done by extracting information such as SSID and MAC address relevant to a specific AP. Technically speaking, RSS is measured by a WLAN receiver after capturing the whole broadcast packet based on modulation and packet length obtained from the PHY header.

More specifically, the MPDU for the beacon frame contains a Frame control (see Figure 2) field which specifies whether the packet being captured is of the management type, and beacon frame subtype. This type and subtype should read 0 for management, and 8 for beacon frame, respectively [15]. Once the packet has been completely received and demodulated, the WLAN receiver reports the RSS value at that moment, thus associating the measurement with that packet. As a summary, WLAN receivers sense the wireless medium (WM) and begin to search for a potential WLAN frame by correlating with the known PHY preamble, i.e. the SYNC. After this, the PHY header describes what is contained in the MPDU and how to demodulate it. Contents on the PHY header are the SIGNAL and LENGTH fields which describe the modulation type and the total length of the PPDU, respectively. Once the MPDU is finally demodulated and information extracted from the fields, the WLAN receiver reports the RSS with its associated information such as SSID, MAC address, etc. In this paper, our intention is to collect channel information not available by commercial WLAN cards and can be extracted by equalizing with the PHY preamble SYNC training sequence. The SDR implementation is detailed in the next section.

**3 PROPOSED WLAN RECEIVER ARCHITECTURE**

The development SDR is implemented by Software Communications and Navigation Systems (SCNS) Laboratory at the University of Texas at San Antonio (UTSA). SDR is defined as a radio in which some or all of its physical layer functions are defined by software [17]. Having this definition, we implemented an 802.11b WLAN receiver in a software radio environment by using LabVIEW (LV) as a software prototyping platform. LV is a platform which uses a visual programming language [18] from National Instruments and it also helps as the interface between the host PC and a front-end, namely the NI-USRP 2932 from National Instruments. We focus on capturing DSSS beacon frames, as many other WLAN functionalities are not required. Figure 3 shows an overview of the 802.11b WLAN SDR receiver. The implementation uses a combination of built-in function blocks along with specific baseband functions that have been compiled in C++ as dynamic link libraries (DLLs) that can be integrated into the software platform for fast prototyping. Other acceleration features that are inherent from LV are also utilized. The next subsections will describe its overall functionality.

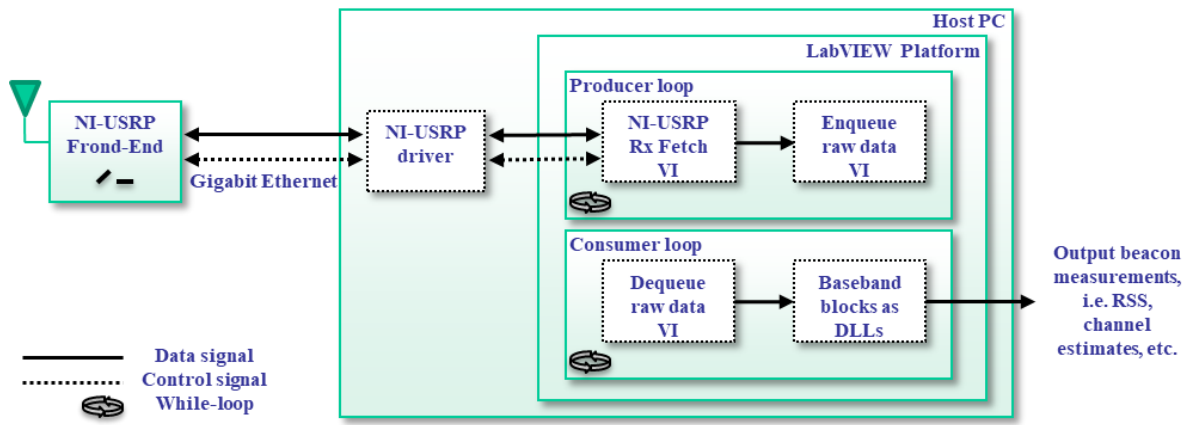

**Figure 3. 802.11b WLAN SDR receiver architecture.**

**3.1 LabVIEW acceleration features**

LV is a development environment based on blocks and virtual instruments (VIs) which can be visually programmed and attached as a main or sub-VI. There is always a main VI where all the upper layer functionality is held such as data flow and main program execution. Based on the characteristic visual programming, LV becomes effective for fast prototyping and data flow handling, thus ideal for the proposed WLAN receiver.

LV has two main components in its environment: a front panel, which acts as a graphical user interface (GUI) editor where built-in visualizations and controls are readily available, and a block diagram, where the actual visual programming occurs via wires, sub-VIs, built-in function blocks such as fast Fourier transforms (FFTs), and loop structures such as while and for loops. The programming flow occurs logically from left to right which allows parallelism. Some of the acceleration features that are exploited from LV to obtain a real-time operation of the WLAN receiver can be seen in Figure 4. These acceleration features were used to develop common WLAN baseband functions such as carrier wipe-off, resampling, and FFT-based correlators for peak packet detection, among others.

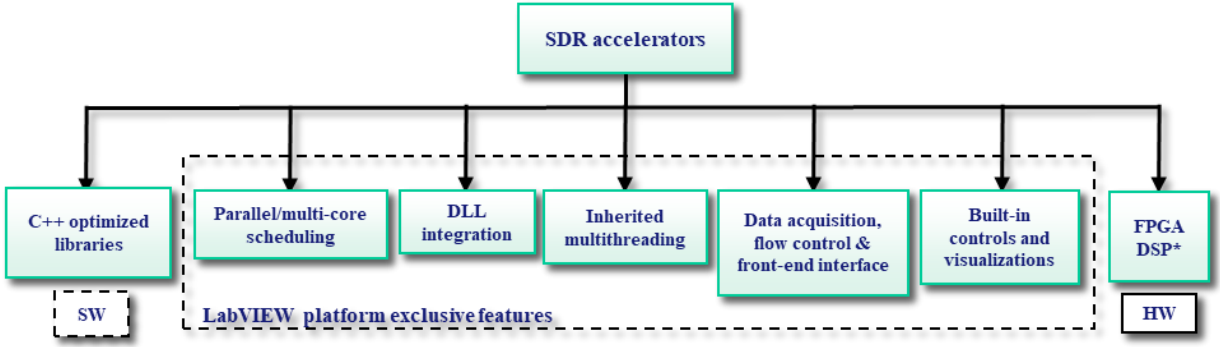

*Optional FPGA blocks have been implemented in previous work [11] but are not used in this SDR implementation.

**Figure 4. Acceleration features used in the proposed WLAN SDR receiver.**

LV has characteristic functionalities that effortlessly assign parallel tasks to multi-core processors by taking advantage of the visual block diagram programming that executes from left to right, and in any order as long as the blocks are independent from each other. Program parallelism achieves performance gains of around 30 percent [9], [10], by concurrently running several independent block diagram paths which are seamless to the developer since LV compiler takes care of scheduling these processes. The compiler also recognizes the host PCs capabilities in terms of cores and threads available. The SDR receiver is strongly tied to LV visual programming for real-time processing and parallelization due to this feature. Said acceleration features have been used in other receiver environments for fast prototyping as in [9]-[11]. Although [11] has even implemented field-programmable gate array (FPGA) hardware accelerators in LV environment, this specific receiver did not had to use such accelerators and still achieved real-time operation.

Complex algorithm development may become overwhelming when too many LV built-in blocks are required to be wired for common baseband modules. Complex algorithms are many times more suited for C/C++ coding, therefore can be found in existing libraries in the form of an application programming interface (API). LabVIEW provides DLL integration by a built-in Call Library Function Node which is used to call functions available in DLL libraries. By coding common WLAN functions such as correlators, carrier wipe-off, look-up tables (LUTs), etc., and compiling them as DLLs, it is easier to concentrate on the isolated algorithm, leaving data flow control and other aspects such as data acquisition and front-end interface to LV. This increases the modularity and adaptation capabilities of the SDR receiver for fast prototyping. Optimized C/C++ libraries feature matching internal architecture for specific resource utilization based on host PC. As much as two-fold acceleration was achieved on proposed SDR receiver for FFT routines by implementing optimization libraries such as FFTW [19]. These optimized libraries have internal functions that try to exploit as much as possible the host PCs architecture and processor capabilities. These libraries can, for instance, apply built-in single-input multiple-data (SIMD) functions to the FFT routines, or schedule computations for multi-threading execution based on number of processor cores in the host PC.

### 3.2 Main producer-consumer loop functionality

LV uses native NI-USRP drivers to interface with the chosen front-end (NI-USRP 2932). The driver sends control signals to the front-end and fetches data in a pre-configured size of samples and sends them to the main LV VI, which eventually passes these samples to the DLL baseband functions. This interface is easily handled by LV with built-in NI-USRP configuration and sample fetching VI blocks which configure the front-end and requests blocks of samples. Most of the implemented features are exploited from LabVIEW platform built-in features.

A common application design architecture called the "producer-consumer loop" can be seen in Figure 3 [9], [10]. Based on this loop, LV can handle a real-time continuous operation by acquiring data on the producer loop in a high-priority uninterrupted manner and sending it to a data queue which allows the received raw samples to be stored in memory as they come. After this, the consumer loop dequeues the data from a first-input-first-output (FIFO) buffer and sends it to the baseband DLL blocks. Both producer and consumer loops are basically a while-loop structure, each running indefinitely and in parallel. These loops operate continuously until the user halts receiver execution or the amount of captured packets is achieved. The internal queue or buffer utilized between this loop-pair is automatically handled by LV in terms of memory allocation, thus occurring in the background. These loop structures (i.e., producer-consumer loop) uses inherited multi-threading which translates to seamless operation of the front-end interface with LabVIEW and therefore dataflow can be easily achieved and controlled. It uses queues that are automatically handled by the program, which can be seen as FIFO buffers fed into the C++ DLLs for raw data post-processing. This structure can be seen as the skeleton that holds together the main flow of samples and operations. After the processing of the signal, if an 802.11b DSSS beacon is detected then it is written to a file, along with its RSS value and channel estimates.

### 3.3 Baseband blocks

We specifically implemented the following digital communication blocks on the receiver: resampler, match filter, coarse frequency estimation and correction [20], code phase synchronization, Wiener-based filter equalizer [21], Barker code de-spreader, demodulator, and descrambler. All these digital communication blocks' descriptions were taken from the IEEE 802.11b standard references [15]. Figure 5 shows an actual LV block diagram of said modules and is executed after a beacon frame packet is detected, i.e. as a post-processing routine.

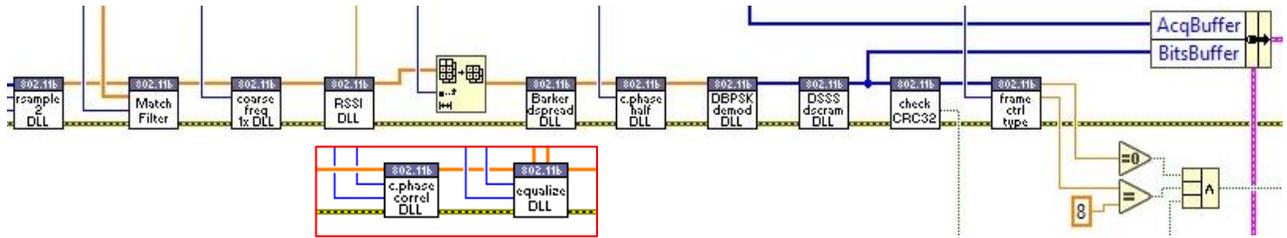

**Figure 5. Baseband modules as sub-VIs that are executed after a beacon frame detection.**

As seen in Figure 5, the first module is a resampler. The NI-USRP 2932 front-end can only configure sampling frequencies that are multiples of 5 MHz, thus the selected sampling rate is 25 MHz which is therefore resamples to 22 MHz for compatibility with DSSS signals [15]. The match filter uses pre-configured filter coefficients that are fed onto a LV built-in filter block, thus DLL was not implemented in this module. The coarse frequency block performs a coarse frequency detection based on [20] and uses a basic carrier wipe-off with the option of using either quantized LUT carriers for faster execution, or full sinusoids. The RSSI block computes the signal strength of the currently detected frame based on synchronization boundaries and by using the PHY header LENGTH to obtain the total number of samples in the packet. A Barker despread module followed by a DBPSK demodulator, and descramblers are seen next, and are based on the standard's specifications [15]. Finally a CRC32 check module is used to verify packet errors and is also based on the standard. Finally, the frame control header is checked to see if the frame type is zero and subtype is 8, as seen in Figure 5. This corresponds to a beacon frame format. Between RSSI DLL and Barker despread DLL there are two other modules that are executed in the packet detection phase occurring before (not seen in this figure). They correspond to a phase correction and an equalization using a Wiener-based filter [21] by using the SYNC sequence. As mentioned in Section 2, the known SYNC training sequence is used against the received synchronized SYNC signal to detect a packet and to further equalize it and collect the channel estimates based on the pre-configured number of channel taps. The Wiener-Hopf equation is used as in [21] to obtain these channel estimates as complex coefficient taps.

### 3.4 Sleep-mode for real-time operation

As of the current version, the SDR is not able to operate in real-time fully. Since the most demanding operation is the packet detection phase which consists of three highly demanding blocks: resampling, coarse frequency estimation and wipe-off, and correlation detection for the SYNC sequence, this phase makes the receiver run at three times slower than real-time. Other accelerators not mentioned in Section 3.1 such as SIMD-based correlators [9], [10], have been known to accelerate significantly the receiver. But since a continuous operation of the receiver is not required, a sleep-mode has been implemented as follows. As seen in [16], WLAN beacons are transmitted periodically based on time units (TUs). A single TU consists of fixed 1024 microseconds, and typical APs transmit beacons with a beacon interval of 100 TUs, which correspond to 102.4 milliseconds. Based on this information and to avoid the receiver to operate in packet detection mode in continuous mode, thus lagging the consumer loop, the receiver is put to "sleep" after detecting a beacon frame packet. This is done by many commercial NICs to save power on the devices and therefore it has been imitated in the SDR implementation. Additionally, we only require beacon frame collection measurement for our research purposes. After sleep-mode is implemented, the receiver is now able to operate in real-time and has achieved around 9.5 packets per second, which is close to the theoretical maximum of 9.76 packet per second as mentioned in [16].

**3.5 Receiver front panel, inputs, and outputs**

The receiver's main front-panel can be seen in Figure 6 (left). The current version when writing this paper is v4.1 as it has been optimized further with functionality such as sleep mode, and quantized carrier options. On the left side, the main configuration section can be seen with basic and advanced options. The Device IP Address corresponds to the front-end device address, along with the receiver gain, and sampling rate. The WLAN Channel corresponds to channels seen in Figure 6 from Section 2 [15]. The output file appends a number to correspond to the location file that is set when surveying is performed and the receiver stops when capturing specified beacons (-1 for unlimited). The receiver is capable of simulation mode; that is, process a binary offline recorded file in INT16 baseband interleaved I-and-Q sample format. As per the advanced configuration, the buffer size is typically left as is, as it corresponds to a sliding window used to detect incoming packets and corresponds to the size of a SYNC sequence. The correlation peak threshold typically ranges between 700-1000 and its value is not scaled but should be understood as 70 to 100 percent threshold. A quantized LUT option can be used for carrier wipe-off and for phase corrections. An optional match filter can be used. Sleep mode can be enabled along with the sleep time after it detects a packet, and finally the equalizer type can be configured for chip-level or symbol-level equalization. Whether symbol or chip level is chosen, the Barker despread block is placed accordingly. Typically chip-level equalization is used, as more channel information can be found on the received chip-level SYNC signal.

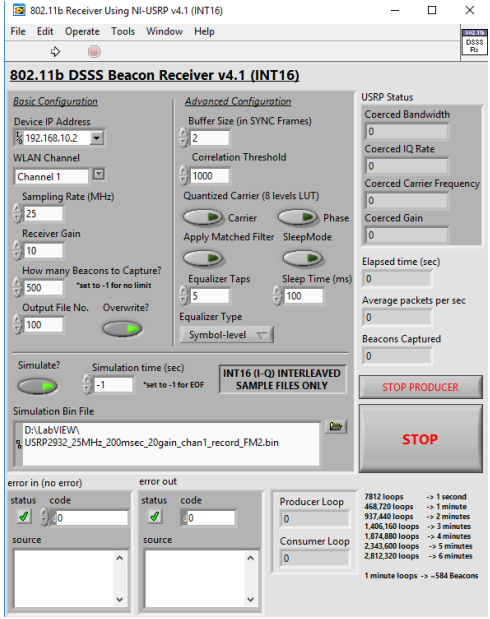

**Figure 6. 802.11b DSSS SDR receiver front-panel (left) and output sample (right).**

As outputs, for each beacon captured in a running session, the receiver provides a timestamp, the SSID of the AP, a MAC-ID from the specified network broadcasting the beacon, WLAN channel number, RSS indicator value in dBm scale which typically ranges from 0 to -100 dBm and is calibrated to the receiver gain configuration, and all channel estimate taps as complex numbers separated by its real and imaginary parts on each column of each tap, non-normalized. For this experiment, total of 5 taps was chosen. Figure 6 (right) shows an example of said outputs.

**4 TESTING METHODOLOGY**

This section describes the proposed indoor positioning system under testing to validate channel estimates as potential signature inputs for improved classification. We have chosen a support vector machine (SVM) as seen in [14] to be used as a classifier and we use DSSS frames to extract RSS and channel estimates to build a radio-map and a localization system. The SVM uses a training phase which is used to build a training model. This training model is then used to predict the user location based on neural network functionality [14]. Since the SVM takes real data as inputs, both the RSS values along with the channel estimate complex values were normalized. Also, channel estimate magnitudes were used as inputs. For SVM, each input that represents a fingerprint for a given location is called a feature. Therefore, the RSS measurements are one feature, and each channel estimate tap is also considered a single feature.

## 4.1 DSSS frames for backward compatibility

Although there are many works in literature related to IPS and channel information such as channel state information (CSI) found in OFDM frames from 802.11n standard [22], this paper focuses on a single standard for beacon capturing. The 802.11b DSSS channel estimates obtained from the previously described SDR are selected for the proposed IPS. We attempt to use said channel estimates because of the backward compatibility of DSSS frames over OFDM frames and other standards. WLAN infrastructure is abundant and it is a reasonable assumption to rely on 802.11b frames to depict a WLAN-scarce environment. Additionally, we test a single AP in our IPS to simulate said scarce WLAN environment with few available APs along with a common cluttered scenario. This cluttered environment should complement to the uniqueness of the fingerprint and the channel. Finally, the novelty in fast prototyping SDRs adds the possibility of extracting any measurement from physical layer baseband signals such as the channel estimate coefficients from DSSS legacy frames. Previous work on [14] has shown an evident improvement in accuracy when using channel information and we aim to improve said results.

## 4.2 Testing scenario

We have performed data collection at the University of Texas at San Antonio (UTSA) at the SCNS lab. We have chosen a granularity of 2 feet (60.96 cm) between reference points (RPs) and a total of 69 RPs were selected in an indoor laboratory depicting a cluttered environment with desks, chairs, lockers, etc. Figure 7 shows the indoor configuration along with the single AP location set to broadcast SSID "TEST-B" on channel 1 in legacy mode. The total area is around 60 sq. m., so the sides are roughly 10 m x 6 m.

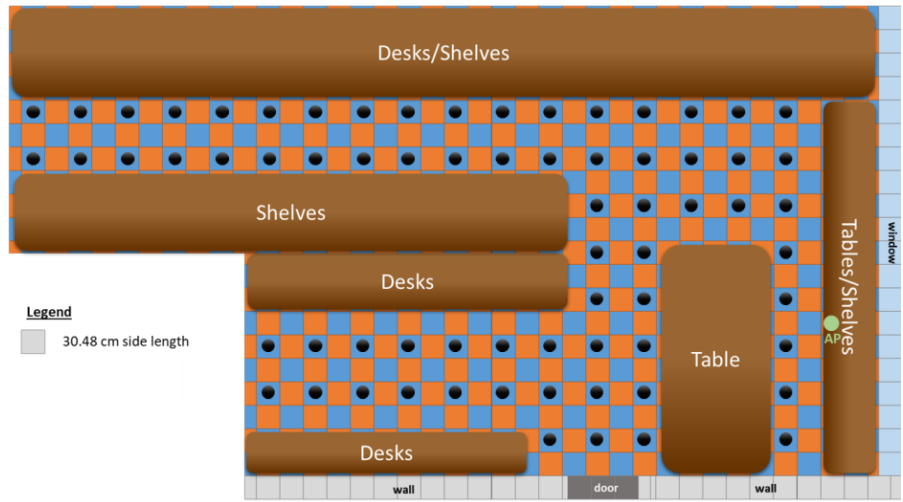

**Figure 7. SCNS Laboratory environment with 69 RPs for proposed IPS.**

For each RP we collected 600 beacon frame measurements, totaling 41,400 measurements for the experiment. Measurements for each RP were chosen randomly to be used as follows: 500 for offline stage SVM training, and 100 for online stage navigation. The survey was conducted in a normal weekday where there was movement, and a tray with the USRP front-end and an ASUS ROG GL552VW laptop with a quad-core Intel i7-6700HQ processor, 32 GB RAM, and Windows 10 with the proposed SDR v4.1 was used. Since we have RSS and channel estimates (in magnitudes), we computed three testing scenarios for a single AP: RSS-only, channel estimates (magnitudes) only, and both. This was done as an attempt to show potential accuracy improvement of said features increasingly. Therefore, the SVM was fed with 1 feature, 5 features, and 6 features, for each scenario respectively.

## 5 RESULTS

For accuracy testing, we computed two statistics, namely prediction accuracy and distance accuracy. For prediction accuracy we used all 100 testing measurement for each RP. The SVM outputs a predicted RP for each of those measurements. Therefore, we computed a simple prediction error statistic based on number of correctly guessed test measurements out of those 100 for each RP. For distance accuracy, we computed actual distance error based in meters from those predictions and computed a mean distance error curve based on a cumulative distribution function (CDF) to compute 50[th] and 90[th] percentile error for said scenarios.

### 5.1 Results for 1 AP and 2 ft. granularity

We computed prediction and distance accuracy for a single AP in the laboratory using previously mentioned SVM classification algorithms from built-in MATLAB environment and 2 ft. granularity and 5 channel taps as magnitudes, thus totaling 6 features as inputs. Figure 8 (top-left) shows prediction accuracy for all three scenarios: RSS-only, channel magnitudes, and both, and (right) distance accuracy CDF plot results along with 50[th] and 90[th] percentiles (bottom-left).

| Prediction Accuracy | Percentage |
|---|---|
| RSS-only | 29.64% |
| Channel estimates | 74.93% |
| RSS + channel estimates | **81.90%** |

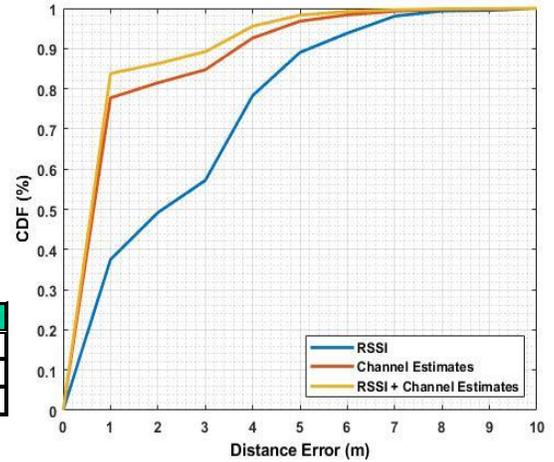

| Distance Error | 50[th] Percentile | 90[th] Percentile | Mean | Std. dev. |
|---|---|---|---|---|
| RSS-only | 2.10 m | 5.20 m | 2.34 m | 2.15 m |
| Channel estimates (5) | 0.64 m | 3.67 m | 0.82 m | 1.65 m |
| RSS + channel estimates | 0.59 m | 3.13 m | 0.57 m | 1.38 m |

**Figure 8. Prediction accuracy results (top-left), CDF distance error plot (right), and CDF error percentiles (bottom-left) for 1 AP with 2 ft. granularity scenario.**

An evident improvement is seen in the CDF plot for distance error for the case of channel estimates and channel estimates with RSS over RSS-only case, as features for SVM classification for a single AP and 2 ft. granularity scenario. From the perspective of prediction accuracy, a gain of 45% in accuracy is achieved when using only channel estimates, and a gain of 52.3% in accuracy is seen when using both RSS and channel estimates in the SVM-based classification. As for distance error, a mean distance error of 2.34 m is seen for 1 AP with RSS-only classification, a mean distance error of 0.82 m is seen with only channel estimates, and 0.57 m mean error for RSS and channel estimates. Their 50[th] percentiles see an improvement of 1.46 m and 1.51 m for channel estimates, and RSS with channel estimates over RSS only, respectively.

## CONCLUSION

This paper presented an IPS which utilizes channel estimates from 802.11b DSSS beacon frames as inputs for an SVM classification algorithm. Said channel estimates are potentially characterizing of each indoor location as a unique fingerprint. To extract channel estimate measurements, commercial WLAN receivers do not provide baseband information. Therefore, a novel fast prototyping WLAN beacon receiver was presented and used to extract said channel estimates. The proposed SDR used a LV-based prototyping platform with several presented acceleration features that improved said receiver for real-time operation. A testing scenario of 69 RPs in SCNS lab was proposed, with 600 measurements taken at each RP for training and testing of the SVM classification system. A single AP was used to test the RSS and channel estimates scenario for potential accuracy improvements. A clear gain in the IPS accuracy results of up to 52% was observed when using RSS and channel estimates for a single AP and 2 ft. granularity in said indoor laboratory location. The proposed IPS generated a mean distance error of 0.57 m for the scenario of RSS and channel estimate measurements, thus proving a potential fingerprinting unique scheme when the channel information is used for classification systems. Future work would suggest an increased number of channel taps, e.g., 10 or 15 taps, which would translate to more descriptive measurements of cluttered indoor environments for online navigation. Other classification algorithms such as k-nearest neighbor and Bayesian classification [12] is also intended for comparison purposes.

**REFERENCES**

[1] GPS.gov, "The Global Positioning System," *GPS.gov: GPS Overview*. N.p., 11 Feb. 2014. Web. http://www.gps.gov/systems/gps/ (2014).

[2] Liu, H., Darabi, H., Janerjee, P., Liu, J., "Survey of Wireless Indoor Positioning Techniques and Systems," on *IEEE Transactions on Systems, Man, and Cybernetics-Part C: Applications and Reviews*, Vol. 37, No. 6, November 2007.

[3] Yassin, M., Rachid, E., Nasrallah, R., "Performance Comparison of Positioning Techniques in Wi-Fi Networks," *IEEE 2014 10th International Conference on Innovations in Information Technology (INNOVATIONS)*, 2014, pp. 75-79.

[4] Wu, D., Xu, Y., Ma, L., "Research on RSS based Indoor Location Method," in *2009 Pacific-Asia Conference on Knowledge Engineering and Software Engineering, Pacific-Asia Conference on Knowledge Engineering and Software Engineering, 2009. KESE '09*, 2009, pp. 205-208.

[5] Dawes, B. and Chin, K., "A comparison of deterministic and probabilistic methods for indoor localization," *The Journal of Systems and Software,* Vol. 84, 2011, pp. 442–451.

[6] Bahl, P., and Padmanabhan, V., "RADAR: an in-building RF-based user location and tracking system," in *Proceedings of the 19th Annual Joint Conference of the IEEE Computer and Communications Societies*, Vol. 2, 2000, pp. 775–784.

[7] Lin, T.-N., Lin, P.-C., "Performance comparison of indoor positioning techniques based on location fingerprinting in wireless networks," *IEEE 2005 International Conference on Wireless Networks, Communications and Mobile Computing (Volume:2)*, Vol. 2, 2005, pp. 1569-1574.

[8] He, S. and Chan, S.-H. G., "Wi-Fi Fingerprint-Based Indoor Positioning: Recent Advances and Comparisons," *IEEE Communication Surveys and Tutorials.,* Vol. 18, No. 1, 2016, pp. 466-490

[9] Schmidt, E., Akopian, D. and Pack, D. J., "Development of a real-time software-defined GPS receiver in a LabVIEW-based instrumentation environment," *IEEE Transactions on Instrumentation and Measurements*, Vol. 67, No. 9, Sept. 2018, pp. 2082–2096.

[10] Schmidt, E. and Akopian, D., "Exploiting acceleration features of LabVIEW platform for real-time GNSS software receiver optimization," in *Proc. 30th Int. Tech. Meeting Satellite Division Inst. Navigat. (ION GNSS), Portland, OR, USA, Sep. 2017*, pp. 3694–3709.

[11] Soghoyan, A., Suleiman, A. and Akopian, D., "A Development and Testing Instrumentation for GPS Software Defined Radio with Fast FPGA Prototyping Support," *IEEE Transactions on Instrumentation and Measurements*, Vol. 63, No. 8, 2014, pp. 2001-2012.

[12] Kushki, A., Plataniotis, K. N., Venetsanopoulos, A. N., "Kernel-Based Positioning in Wireless Local Area Networks," *IEEE Transactions on Mobile Computing,* Vol. 6, No. 6, Jun. 2007, pp. 689-705.

[13] Del Mundo, L.B., Ansay, R.L.D., Festin, C.A.M., Ocampo, R.M., "A Comparison of Wireless Fidelity (Wi-Fi) Fingerprinting Techniques," *IEEE International Conference on ICT Convergence (ICTC)*, 2011, pp. 20-25.

[14] Schmidt, E., Akopian, D., "Indoor Positioning System Using WLAN Channel Estimates as Fingerprints for Mobile Devices," in *Proc. of SPIE-IS&T Electronic Imaging*, Vol. 9411, 2015.

[15] IEEE Computer Society, "IEEE Standard for Information technology--Telecommunications and information exchange between systems Local and metropolitan area networks--Specific requirements Part 11: Wireless LAN Medium Access Control (MAC) and Physical Layer (PHY) Specifications," 1-2793 (2012).

[16] Schmidt, E., Mohahmed, M.A. and Akopian, D., "A Performance Study of a Fast-Rate WLAN Fingerprint Measurement Collection Method," *IEEE Transactions on Instrumentation and Measurements*, Vol. 67, No. 10, Oct. 2018, pp. 2273-2281.

[17] National Instruments. Software Defined Radio. [Online]. Available: http://www.ni.com/en-us/shop/select/software-defined-radio-device. [Accessed: August 3, 2017].

[18] National Instruments, "What Is LabVIEW?," NI Newsletters, Publish Date: Aug 16, 2013. Retrieved from http://www.ni.com/newsletter/51141 (2015).

[19] Frigo, M. and Johnson, S. G., "The Design and Implementation of FFTW3," *Proceedings of the IEEE*, Vol. 93, No. 2, 2005, pp. 216-231.

[20] Luise, M. and Reggiannini, R., "Carrier frequency recovery in all-digital modems for burst-mode transmissions," *IEEE Transactions on Communications*, Vol. 43, No. 2-3-4, Feb.-March-Apr. 1995, pp. 1169-1178.

[21] Mehrpouyan, H., "Channel equalizer design based on wiener filter and least mean square algorithms," in *Submitted to EE517 at RMC*, pp. 1–7, 2009.
[22] Wang, X., Gao, L., Mao S. and Pandey, S., "CSI-Based Fingerprinting for Indoor Localization: A Deep Learning Approach," *IEEE Trans. Veh. Technol.*, Vol. 66, No. 1, 2017, pp. 763-776.